\documentclass[onecolumn,showpacs,amsfonts,aps,prc,nofootinbib,floatfix,%
superscriptaddress]{revtex4}

\usepackage{amsmath}
\usepackage{bm}
\usepackage{graphicx}

\newcommand{\tg}{\tan}

\usepackage{epsfig}
\newcommand{\beq}{\begin{equation}}
\newcommand{\eeq}{\end{equation}}
\newcommand{\bea}{\vspace{0.25cm}\begin{eqnarray}}
\newcommand{\eea}{\end{eqnarray}}

\newcommand{\ta}{\mbox{{\boldmath
$\tau$}}}

\newcommand{\ro}{\mbox{{\boldmath
$\rho$}}}

\newcommand{\qb}{\mbox{{\bf
q}}}

\newcommand{\ub}{{{\bf u}}}

\newcommand{\Rb}{{{\bf R}}}

\newcommand{\pt}{\mbox{{\bf
p}}_\perp}

\def\lsim{\mathrel{\rlap{\lower4pt\hbox{\hskip1pt$\sim$}}
    \raise1pt\hbox{$<$}}}         
\def\gsim{\mathrel{\rlap{\lower4pt\hbox{\hskip1pt$\sim$}}
    \raise1pt\hbox{$>$}}}         

\newcommand{\landau}{L.D.~Landau Institute for Theoretical Physics,
        GSP-1, 117940, Kosygina Str. 2, 117334 Moscow, Russia}

\begin{document}


\title{Radiative $p_{\perp}$-broadening of fast partons in an expanding quark-gluon plasma
}
\date{\today}

\author{B.G.~Zakharov}\affiliation{\landau}

\begin{abstract}
We study contribution  of  radiative  processes  to   
$p_{\perp}$-broadening  of  fast  partons  in  an expanding   quark-gluon
plasma.
It is shown that the radiative correction to 
$\langle p_{\perp}^2\rangle$ for the QGP produced in $AA$-collisions at 
RHIC and LHC may be negative, and comparable in absolute value with the
non-radiative contribution.
We have found that the QGP expansion enhances the radiative suppression of
$p_\perp$-broadening as compared to the static medium. 
\end{abstract}
%

\maketitle

\section{Introduction}
One of the major signals of the quark-gluon plasma (QGP) formation
in heavy ion collisions at RHIC and LHC is the strong suppression
of particle spectra at high transverse momenta.
It is believed
that this effect (usually called the jet quenching) is a consequence
of parton energy loss in the QGP, which softens the jet fragmentation
functions.
The parton energy loss is dominated by
the radiative mechanism through induced gluon emission
\cite{GW,BDMPS1,BDMPS2,LCPI1,W1,GLV1,AMY}.
The induced gluon emission is caused by multiple scattering
of fast partons in the QGP.
The  induced gluon spectrum can  be  expressed  via  the Green function  of
a  2D Schr\"odinger equation with an imaginary potential \cite{LCPI1,BDMPS1},
which  is proportional to  the  product $n\sigma_{q\bar{q}}(\rho)$, where
$n$ is the QGP number density and $\sigma_{q\bar{q}}(\rho)$ is
the dipole cross section  of scattering of
a $q\bar{q}$ pair off the QGP constituent (here, 
$\rho$ is the size of the $q\bar{q}$-pair).
In the  quadratic approximation $\sigma_{q\bar{q}}(\rho)\approx C\rho^2$,
the Hamiltonian of the Schr\"odinger equation takes the oscillator form
with  a  complex  frequency. The square of
the frequency is proportional to the transport coefficient $\hat{q}$
\cite{BDMPS1,BDMPS2} defined by the
relation   $\hat{q}=2Cn$.

Besides the modification of the longitudinal jet structure due to
the parton energy loss,
multiple parton scattering in the QGP can also
modify the transverse structure of the jet.
For a single fast parton
the mean squared transverse momentum (relative to
its initial velocity) in a uniform medium is given by \cite{BDMPS2}
\beq
\langle p_{\perp}^2\rangle=\hat{q}L\,,
\label{eq:10}
\eeq 
where $L$ is the path length in the medium.
One could expect that the $p_\perp$-broadening  of  the leading parton in
the jet should lead to an increase of azimuthal jet decorrelation in
the di-jet events (or in decorrelation of a photon and
the  jet  in  the  photon-jet  events)  in  $AA$-collisions
\cite{Mueller_dijet}.
For  a better understanding  of the in-medium jet evolution,  it  would  be
interesting to compare the values of $\hat{q}$   extracted from
the $R_{AA}$ data with that obtained from the jet $p_\perp$-broadening.
However, due to a considerable background from 
the azimuthal jet decorrelation in $pp$ collisions related to
the  Sudakov  formfactors \cite{Mueller_dijet},
experimental  detection  of  the jet $p_\perp$-broadening  is a difficult
problem. This background is especially large for LHC energies.
However, even at RHIC in Au+Au collisions at $\sqrt{s}=0.2$ TeV,
where the effect of Sudakov formfactors are weaker,
the STAR Collaboration has not detected a statistically significant
effect of the jet deflection in the QGP \cite{STAR1}.
The first data from ALICE \cite{ALICE_hjet} for Pb+Pb collisions at
$\sqrt{s}=5.02$ TeV also do not allow to draw a definite conclusion on
the jet $p_\perp$-broadening. But the new preliminary ALICE results
\cite{ALICE_hjet2} indicate
that the jet $p_\perp$ distribution in 5.02 TeV Pb+Pb collisions
may be somewhat narrower than that in $pp$ collisions.
The situation with detecting  the jet $p_\perp$-broadening
can become better after improving the accuracy of the
data \cite{Gyulassy_dijet,PJacobs}.

On the theoretical side, it would be interesting to understand the role
of the radiative contribution to $p_\perp$-broadening.
The radiative correction to $p_\perp$-broadening can come from the real and virtual
induced gluon emission \cite{Wu,Mueller_pt,Blaizot_pt}.
It was expected that, due to smallness 
of the formation length for dominating soft gluon emission,
this effect can be viewed as a local renormalization of $\hat{q}$.
In \cite{Mueller_pt}, within the oscillator approximation in
the soft gluon limit, it was found that the radiative contribution  to the
mean $p_\perp^2$,   
$\langle p_{\perp}^2\rangle_r$, in a homogeneous  QGP   has  a double
logarithmic form
\beq
\langle p_{\perp}^2\rangle_r\sim \frac{\alpha_sN_c\hat{q}L}{\pi}\ln^2(L/l_0)\,,
\label{eq:20}
\eeq
 where $l_0$  is the size on the order of the Debye radius in
 the QGP. For  central $AA$-collisions ($L\sim 5$ fm)
the radiative contribution to $\langle p_{\perp}^2\rangle$
turns out to be comparable  with that from ordinary multiple
scattering \cite{Mueller_pt}.
In \cite{Mueller_pt} the authors used the light-cone path integral
(LCPI) formalism developed in \cite{LCPI1}.
The case of the transverse spectra has been addressed within the LCPI technique
in our earlier work \cite{LCPI_PT} (see also \cite{BSZ,W1}).
In the formulation of \cite{LCPI_PT}, for $a\to bc$ transition
the distribution of the particle $b$ in the Feynman variable $x$ and
the transverse momentum  is described by the
diagram of Fig.~1a. The contribution from the virtual process
$a\to bc\to a$ to the distribution of the final particle $a$ is described
by the diagram of Fig.~1b. In the case of $a=b$ both the diagrams
contribute to the radiative correction to $p_\perp$-broadening.
The parallel lines for two-body parts in Fig.~1  correspond to the
Glauber factors, that describe the initial and final state interaction for real
and virtual processes, and the three-body part describes dynamics
of the transverse motion of the $bc\bar{a}$-state. 
The analytical expressions for the diagrams of Fig.~1 will be discussed
below. Calculations of \cite{Mueller_pt} correspond to
the diagrams of Fig.~1, but the authors have not accounted for the effect of the
Glauber factors for the initial and final states.
In \cite{Z-pt-JETPL,Z-pt-JETP} (as in \cite{Mueller_pt} for a static medium),
we have addressed the radiative correction to
$p_\perp$-broadening with an accurate treatment of the Glauber factors.
It was found that
the effect of the final state Glauber factors on
$\langle p_{\perp}^2\rangle_r$
vanishes for the sum of the real and virtual diagrams.
However, the initial state Glauber factors give a considerable negative
contribution to $\langle p_{\perp}^2\rangle_r$, and its absolute magnitude
turns out to be bigger than the positive contribution from the diagrams of
Fig.~1 evaluated without the Glauber factors. As a result, the radiative
contribution to the mean $p_\perp^2$ turns out to be negative,
and comparable to the ordinary non-radiative mean $p_\perp^2$ given by
(\ref{eq:10}) (below we denote it by $\langle p_{\perp}^2\rangle_0$).
If this really occurs, then the absence of a signal of the jet deflection in
the data \cite{STAR1, ALICE_hjet} may
be due to a considerable compensation between the radiative and
non-radiative contributions to $p_\perp$-broadening.
To understand better whether or not this scenario is possible  
it is highly desirable to study the radiative
$p_\perp$-broadening for a more realistic model with an expanding QGP.
This is the purpose of the present paper. The case of the expanding QGP
has been addressed previously in \cite{Iancu-pt}. But there
the effect of the Glauber factors has not been accounted for.

For an expanding QGP the transport coefficient decreases with the proper
time $\tau$.  In the Bjorken model \cite{Bjorken} without the transverse
expansion $\hat{q}(\tau)=\hat{q}(\tau_0)(\tau_0/\tau)$ \cite{Baier_q},
where $\tau_0$ is the QGP formation time.
In this case, in the oscillator approximation, the induced gluon spectrum can
be expressed through the Green function for the
oscillator frequency $\propto 1/\sqrt{\tau}$. The induced gluon
emission in the oscillator approximation, in an expanding QGP
has been addressed in \cite{BDMS}.
There it was shown that for the transport coefficient
$\hat{q}(\tau)=\hat{q}(\tau_0)(\tau_0/\tau)^{\alpha}$ the 
resulting total radiative energy loss coincides with that for a static
medium with an effective transport coefficient given by
\beq
\hat{q}_{st}=\frac{2}{L^2}\int_{\tau_0}^{L}
d\tau \tau \hat{q}(\tau)\approx \frac{2}{2-\alpha}\hat{q}(L)\,.
\label{eq:30}
\eeq
In \cite{SW} it was demonstrated by numerical calculations
that for $\alpha\sim 0-1.5$ this scaling law works very well for the
induced gluon spectrum as well.
This means that for decreasing with $\tau$ $\hat{q}(\tau)$
the reduction of the radiation rate in the region
of large $\tau$, where $\hat{q}(\tau)<\hat{q}_{st}$,
is almost compensated by its excess from the region of small
$\tau$, where $\hat{q}(\tau)>\hat{q}_{st}$.
Of course, this scaling law is not valid for the non-radiative
contribution to the mean $p_\perp^2$, which for an expanding
medium reads
\beq
\langle p_{\perp}^2\rangle_{0}=\int_0^L d \tau \hat{q}(\tau)\,.
\label{eq:40}
\eeq
One could expect that it does not hold
for the radiative contribution $\langle p_{\perp}^2\rangle_r$ as well.
Because there is no reason for the delicate
compensation between the regions of large and small $\tau$, even if it
occurs for the gluon spectrum.
However, from the point of view of the jet $p_\perp$-broadening
the most interesting quantity is the ratio $\langle p_{\perp}^2\rangle_r/\langle p_{\perp}^2\rangle_{0}$,
(which characterizes the relative effect of the radiative correction).
For this ratio the violation of the dynamical scaling potentially could be
smaller than that for $\langle p_{\perp}^2\rangle_r$ and $\langle
p_{\perp}^2\rangle_{0}$ separately.
Intuitively, one can expect that for the expanding scenario
the ratio $\langle p_{\perp}^2\rangle_r/\langle p_{\perp}^2\rangle_{0}$
should be negative and bigger, in absolute value, than that for the static
case. Because, the effect of the negative
contribution from the initial parton rescatterings
(which is mostly sensitive to the region of small $\tau$)
should be more pronounced
for the decreasing $\hat{q}(\tau)$.
Our numerical calculations confirm this.

The paper is organized as follows. In Section 2, we
discuss the method for evaluation of the radiative contribution
to the mean $p_\perp^2$ within the LCPI approach.
In  Section  3, we present the results of numerical calculations.
Conclusions  are  contained  in Section 4. 
In appendix, we give formulas necessary for numerical calculations
of $\langle p_{\perp}^2\rangle_r$ in the oscillator approximation.

\begin{figure}[h]
\begin{center}
\includegraphics[height=3.9cm]{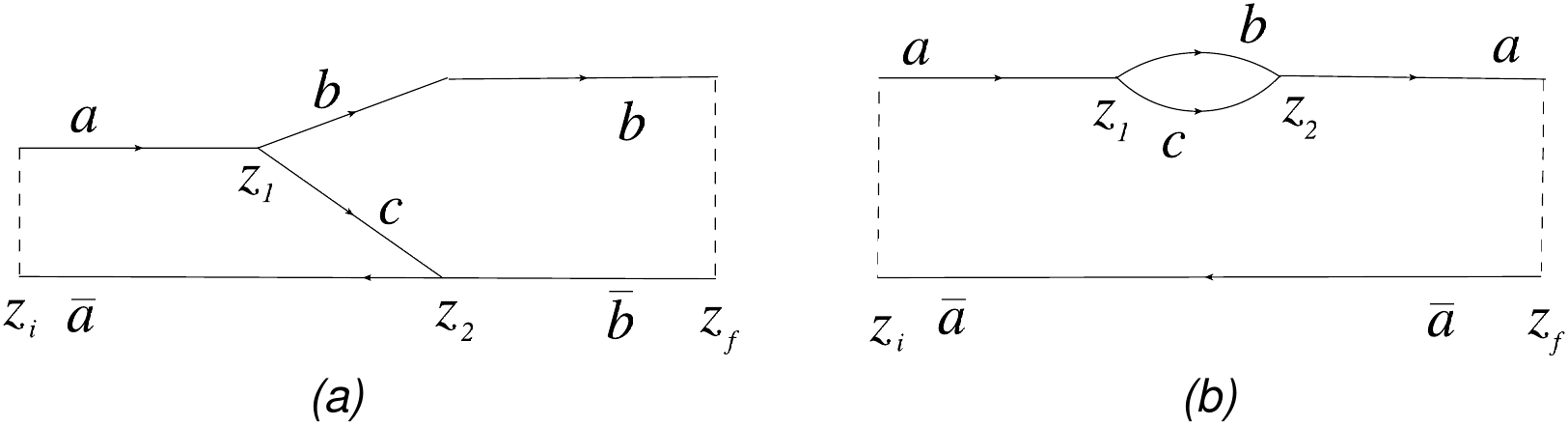}
\end{center}
\caption[.]{
(a)  Diagram  representation  for  the  spectrum of the particle $b$
  for the $a\to bc$ process.
(b)  Diagram  representation  for 
  the radiative correction to the spectrum of the particle $a$
from the virtual process $a\to bc\to a$. There are also 
similar diagrams  with  transposition  of  vertices  between  the
upper and lower parts of diagrams (a) and (b).}
\end{figure}

\section{Theoretical framework for evaluation of $p_\perp$-broadening in LCPI
  approach}
We will consider $p_\perp$-broadening for a fast quark, when
the real process is $q\to q g$ splitting (i.e., $a=b=q$ and $c=g$),
and the virtual one is $q\to qg\to q$.
We assume that the initial quark with energy $E$ is produced  at $z=0$ (we choose the $z$-axis along
the initial quark momentum)  in the QGP of thickness $L$.
The radiative correction to the distribution of the final quark in $x$ and
$p_\perp$ from the real
process $q\to qg$ is described by the diagram of the type shown in Fig.~1a.
And the virtual contribution is described by the diagram of the type of shown
in Fig.~1b.
If one  disregards the collisional  
parton energy  loss,
the total energy
of the two-parton state and the energy of the one-parton  state  
are  identical  at $z\to \infty$.
However, the medium changes the relative  weight  of  
the  one-parton  and  two-parton  states, and
their  transverse  momentum  distributions.
As in \cite{Mueller_pt}, we consider the medium effect on the transverse momentum distribution for
the final quark, which is integrated over its
energy. For  the  virtual contribution  the energy remains  unchanged, but
rescatterings  in  the medium of the intermediate two-particle state differ
from  rescatterings  of  a  single  quark.
The radiative contribution to the mean quark $p_\perp^2$
due to the real and virtual process, 
associated  with  the  interaction with the medium can
be written as \cite{Mueller_pt}
\beq
\langle p_\perp^2\rangle_r=\int dx d\pt \pt^2
\left[\frac{dP}{dxd\pt}+\frac{d \tilde{P}}{dxd\pt}\right]_{ind}\,,
\label{eq:50}
\eeq
where $\frac{d P}{dxd\pt}$   is the distribution  
in  the Feynman  variable  $x=x_q$  and  the  transverse
momentum of the quark for real process $q\to qg$, and
$\frac{d\tilde{P}}{dxd\pt}$ is the distribution 
for the virtual process $q\to qg \to q$. The subscript ``ind'' indicates
that the purely vacuum contribution is subtracted.
For the virtual process, 
$x$ is determined by the Feynman variable of the quark in the intermediate 
$qg$ state. The variable $p_\perp$ in (\ref{eq:50}) for 
the real and virtual terms corresponds to the final quarks.
Of course, formula (\ref{eq:50}) 
can  be  written  in  terms  of  the Feynman  variable  for  the  gluon,  
$x_g=E_g/E$,  which  is connected with 
$x_q$ by the relation $x_q+x_g=1$.

Let us consider first the real splitting.
For $a\to bc$ splitting the distribution on the transverse momentum
and the longitudinal fractional momentum of the particle $b$ 
(which includes both the vacuum and the induced contributions)
can be written in the  form \cite{LCPI_PT} (see also \cite{Z-pt-JETP})
\beq
\frac{dP}{dx_b d\pt}=\frac{1}{(2\pi)^{2}}
\int\!\!
d\ta_f\,\exp(-i\pt\ta_f)F(\ta_f)\,,
\label{eq:60}
\eeq
where
\bea
F(\ta_f)=
2\mbox{Re}\!\!
\left.
\int_{0}^{\infty}\!\!dz_{1}\!\!\int_{z_{1}}^{\infty}\!\!dz_{2}
\Phi_{f}(\ta_f,z_{2})
\hat{g}
{\cal{K}}(\ro_{2},z_{2}|\ro_1,z_{1})
\Phi_{i}(\ta_i,z_{1})\right|_{\ro_2=\ta_f, \ro_1=0}
\,,
\label{eq:70}
\eea
\beq
\Phi_{i}(\ta_i,z_{1})=
\exp\left[-\frac{\sigma_{a\bar{a}}(\ta_i)}{2}
\int_{0}^{z_{1}}\!dz n(z)\right]\,,
\label{eq:80}
\eeq
\beq
\Phi_{f}(\ta_f,z_{2})=
\exp\left[-\frac{\sigma_{b\bar{b}}(\ta_f)}{2}
\int_{z_{2}}^{\infty}\!dz n(z)\right]\,,
\label{eq:90}
\eeq
$\ta_i=x_b\ta_f$, 
$n(z)$ is the number density of the medium,
$\sigma_{a\bar{a}}$ and $\sigma_{b\bar{b}}$ are the dipole cross
sections for the $a\bar{a}$ and $b\bar{b}$ pairs,
$\hat{g}$ is the vertex operator,
${\cal{K}}$ is the Green function for the Hamiltonian
\beq
H=\frac{\qb^2+\epsilon^2}{2M}
-\frac{in(z)\sigma_{\bar{a}bc}(\ta_i,\ro)}{2}\,,
\label{eq:100}
\eeq
where $\qb=-i\partial/\partial \ro$, $M=E_{a}x_bx_c$, 
$\epsilon^2=m_{b}^{2}x_c+m_{c}^{2}x_b-m_{a}^{2}x_bx_c$
 with $x_c=1-x_b$,
and 
$\sigma_{\bar{a}bc}$ is the cross section for the 
three-body $\bar{a}bc$ system. The relative transverse parton coordinates 
for the $\bar{a}bc$ state are given by
$\ro_{b}-\ro_{\bar{a}}
=\ta_i+x_c\ro$,
$\ro_{c}-\ro_{\bar{a}}
=\ta_i-x_b\ro$.
The vertex operator in (\ref{eq:70}) reads
\bea
\hat{g}
=\frac{P^b_a(x_b)g(z_1)g(z_2)}{8\pi M^2}
\frac{\partial}{\partial\ro_1}
\frac{\partial}{\partial\ro_2}
\,,
\label{eq:110}
\eea
where $P^b_a(x_b)$ is the standard $a\to b$  splitting function.
Differentiation with respect to $\ro_1$ and $\ro_2$ on the right-hand side
of (\ref{eq:70}) should be performed
at a fixed $\ta_i$, i.e. for a fixed position of the center mass of
the $bc$ pair.
The Glauber factors $\Phi_i$ and $\Phi_f$ in (\ref{eq:70})
correspond to the parallel   lines in Fig.~1
for the initial ($z<z_1$) and final ($z>z_2$) particles, and  the Green 
function ${\cal K}$ describes evolution of the three-body system in Fig.~1
between $z_1$ and $z_2$.
The factor $2$ in (\ref{eq:30}) accounts for the contribution from the diagram 
that can be obtained by inter-exchange of the vertices
between the upper and lower lines in Fig.~1.

For $q\to qg$ splitting
the three-body cross section reads \cite{NZ_SIGMA3}
\beq
\sigma_{\bar{q}qg}=
\frac{9}{8}\left[\sigma_{q\bar{q}}(\ro_{qg})+\sigma_{q\bar{q}}(\ro_{g\bar{q}})\right]
-\frac{1}{8}\sigma_{q\bar{q}}(\ro_{q\bar{q}})\,,
\label{eq:120}
\eeq 
where
\beq
\sigma_{q\bar{q}}(\rho)=C_{F}C_{T}\int d\qb \alpha_{s}^{2}(q^2)
\frac{[1-\exp(i\qb\ro)]}{(q^{2}+m_{D}^{2})^{2}}
=
\frac{2}{\pi}\int d\qb [1-\exp(i\qb\ro)]\frac{d\sigma}{dq^2}\,,
\label{eq:130}
\eeq
is the dipole cross section for the $q\bar{q}$ system,
$m_D$ is the Debye mass, $C_F$ and $C_T$ are
the color Casimir operators for quark and the QGP constituent,
$d\sigma/dq^2$ is the differential cross section for quark scattering
off the QGP constituent.
The ratio $C(\rho)=\sigma_{q\bar{q}}(\rho)/\rho^2$ is a smooth function of $\rho$.
For the quadratic approximation 
\beq
\sigma_{q\bar{q}}(\rho)=C\rho^2,\,\,\,\,\,\,C=\hat{q}C_F/2C_An\,,
\label{eq:140}
\eeq 
the Hamiltonian (\ref{eq:100}) can be written in 
the oscillator form. For an expanding medium the oscillator frequency depends
on  $z$.

The purely vacuum contribution to $a\to bc$ splitting in (\ref{eq:70}) comes
from the region of large $z_{1,2}$ up to $z_f=\infty$.
For an accurate treatment of the contribution of this region,
 an adiabatically switching off coupling
should be used in the vertex factor (\ref{eq:110}).
To separate the vacuum contribution it is convenient to write
the product 
$\Phi_{f} \hat{g} {\cal{K}}\Phi_{i} $
on the right-hand side of (\ref{eq:70}) 
as (we omit arguments for clarity)
\beq
\Phi_f\hat{g}{\cal K}\Phi_i=\Phi_f\hat{g}({\cal{K}}-{\cal{K}}_v)\Phi_i+
(\Phi_f-1)\hat{g}{\cal K}_v\Phi_i
+{\hat{g}\cal{K}}_v(\Phi_i-1)+\hat{g}{\cal K}_v\,.
\label{eq:150}
\eeq
Here ${\cal K}_v$ is the vacuum Green function, and  the
last term on the right-hand side of (\ref{eq:150})
corresponds to the ordinary vacuum splitting. Its contribution
can be calculated using the adiabatically switching off
coupling $g(z)=g\exp(-\delta z)$
and taking the limit $\delta \to 0$
(see  \cite{Z-pt-JETP} for details). This leads to the vacuum
spectrum
\beq
\frac{dP_{v}}{dx d\pt}=\frac{\alpha_sP_{ba}(x)}{2\pi^2}
\frac{\pt^2}{(\pt^2+\epsilon^2)^2}\,,
\label{eq:160}
\eeq
where $P_{ba}$ is  the  conventional $a\to b$ splitting function.
Note that calculation of the medium dependent contribution to the
spectrum of the last but one term in (\ref{eq:150})
also requires using the adiabatically
switching off coupling. The region of very large $z_{1,2}$ is not important
for the first two terms on the right hand side of (\ref{eq:150}), and they can
be calculated with a $z$-independent coupling.

The distribution
$\frac{d \tilde{P}}{dxd\pt}$
for the virtual process 
$a\to bc \to a$, which is described by the diagram of Fig.~1b,
can be written in the form similar to that for the real process by replacing
$F$ by its virtual counterpart $\tilde{F}$, given by 
\bea
\tilde{F}(\ta_f)=
-2\mbox{Re}\!\!
\int_{0}^{\infty}\!\!dz_{1}\!\!\int_{z_{1}}^{\infty}\!\!dz_{2}
\Phi_{f}(\ta_f,z_{2})
\left.
\hat{g}
\tilde{\cal{K}}(\ro_{2},z_{2}|\ro_1,z_{1})
\Phi_{i}(\ta_i,z_{1})\right|_{\ro_2=\ro_1=0}
\,.
\label{eq:161}
\eea
Except for the opposite sign, the functional form of $\tilde{F}$
is similar to that
for $F$, given by (\ref{eq:70}), but now we have $\ta_i=\ta_f$,
and the Green function (we denote it $\tilde{\cal{K}}$),
for the three-body part between $z_1$ and $z_2$,
should be calculated at $\ro_1=\ro_2=0$.
The change of the sign in (\ref{eq:161}) as compared
to (\ref{eq:70}) occurs due to the fact that for
the virtual process both the vertices, for parton splitting and merging,
belong to the amplitude (upper part of the graph of Fig.~1b),
which changes the sign of the vertex operator (\ref{eq:110}).
Note that, similarly to the Green function $\cal{K}$ entering 
the formula (\ref{eq:70}) for $F$, the Green function $\tilde{\cal{K}}$
in (\ref{eq:161}) has a hidden dependence on $\ta_f$ coming from
the $\ta_f$-dependence of its Hamiltonian. 
For $\tilde{\cal{K}}$
the Hamiltonian is also given by (\ref{eq:100}),
but now with $\ta_i=\ta_f$. It is
important that due to different relations
between $\ta_i$ and $\ta_f$ for the real and virtual diagrams,
the initial Glauber factors $\Phi_i(\ta_i)$ in (\ref{eq:70}) and
(\ref{eq:161}) differ as well.

It is evident that $\langle p_\perp^2\rangle_r$ 
can be expressed in terms of the Laplacian of function
$F+\tilde{F}$ with respect to $\ta_f$ 
at $\ta_f=0$ as\footnote{We assume that the factors $F$ and $\tilde{F}$ are
calculated in the quadratic approximation (\ref{eq:140}).
The formula (\ref{eq:170})
should be valid to a logarithmic accuracy beyond the quadratic
approximation as well, if $\ta_f=0$ is replaced by $\ta_f\sim 1/p_{\perp\, max}$.}
\beq
\langle p_\perp^2\rangle_r=
-\int dx \left.[\nabla^2F(\ta_f)+\nabla^2\tilde{F}(\ta_f)]
\right|_{\ta_f=0}\,.
\label{eq:170}
\eeq
We stress that calculation of the Laplacians in (\ref{eq:170})
should be performed treating the functions $F$ and $\tilde{F}$ as
functions of $\ta_f$ only, i.e., using the rigid connections
$\ro_2=\ta_f$, $\ta_i=x\ta_f$  for $F$ and $\ta_i=\ta_f$ for $\tilde{F}$
in calculating the initial  Glauber factors $\Phi_i$ and the Green
functions $\cal{K}$ and $\tilde{\cal{K}}$.
A simple calculation using the identity (\ref{eq:150}), after subtraction
of the purely vacuum contributions, allows to represent
$\langle p_\perp^2\rangle_r$ as a sum
of three terms
\beq
\langle p_\perp^2\rangle_r=I_1+I_2+I_3\,,
\label{eq:180}
\eeq
where $I_i$  are given by
\bea
I_1=-2\int dx\int_{0}^L dz_1
\int_{0}^{\infty} d\Delta z
\mbox{Re}\Big\{\nabla^2\hat{g}
[{\cal{ K}} (\ro_2 , z_2 | \ro_1 , z_1)-
{\cal{ K}}_v (\ro_2 , z_2 | \ro_1 , z_1)]\Big|_{\ro_2=\ta_f,\ro_1=0,\ta_f=0}
\nonumber\\
-\nabla^2\hat{g}
[\tilde{\cal{ K}} (\ro_2 , z_2 | \ro_1 , z_1)-
\tilde{\cal{ K}}_v (\ro_2 , z_2 | \ro_1 , z_1)]\Big|_{\ro_2=\ro_1=0,\ta_f=0}\Big\}\,,
\label{eq:190}
\eea
\bea
I_2=
-2\int dx \int_{0}^L dz_1
\int_{0}^{\infty} d\Delta z
\mbox{Re}\Big\{\hat{g}
[{\cal K}(\ro_2,z_2|\ro_1,z_1)-{\cal K}_v(\ro_2,z_2|\ro_1,z_1)]
\nabla^2\Phi_i(x\ta_f,z_1)\Big|_{\ro_2=\ta_f,\ro_1=0,\ta_f=0}
\nonumber\\
-
\hat{g}[\tilde{{\cal K}}(\ro_2,z_2|\ro_1,z_1)
-\tilde{{\cal K}}_v
(\ro_2,z_2|\ro_1,z_1)]
\nabla^2\Phi_i(\ta_f,z_1)\Big|_{\ro_2=\ro_1=0,\ta_f=0}\Big\}
\nonumber\\
=
-2
\!\int\!dx f(x)\!\!\int_{0}^L\!\!\! dz_1
\langle p_\perp^2(z_1)\rangle_0
\int_{0}^{\infty}d\Delta z
\mbox{Re}\hat{g}\left[
{\cal K}(\ro_2,z_2|\ro_1,z_1)-{\cal K}_v(\ro_2,z_2|\ro_1,z_1)\right]
\Big|_{\ro_2=\ro_1=0,\ta_f=0}\,,
\label{eq:200}
\eea
\bea
I_3=
\int dx
\nabla^2[\Phi_i(\ta_f,L)
-\Phi_i(x\ta_f,L)]\Big|_{\ta_f=0}
\frac{dP_v}{dx}=
-\langle p_\perp^2\rangle_0 \int dx
f(x)\frac{dP_v}{dx}\,,
\label{eq:210}
\eea
Here $\langle p_\perp^2(z_1)\rangle_0=\int_0^{z_1}\hat{q}(z)dz$,
 $f(x)=1-x^2$, $\Delta z=z_2-z_1$, and 
$dP_v/dx$  is the $x$-spectrum for vacuum given by
\beq
\frac{dP_v}{dx}=\int d\pt
\frac{dP_v}{dx d\pt}\,.
\label{eq:220}
\eeq
In the expressions for $I_{2,3}$, we have used that
\bea
\hat{g}
[{\cal K}(\ro_2,z_2|\ro_1,z_1)-{\cal K}_v(\ro_2,z_2|\ro_1,z_1)]\Big|_{\ro_2=\ta_f,\ro_1=0,\ta_f=0}=
\hat{g}[\tilde{{\cal K}}(\ro_2,z_2|\ro_1,z_1)
-\tilde{{\cal K}}_v
(\ro_2,z_2|\ro_1,z_1)]\Big|_{\ro_2=\ro_1=0,\ta_f=0}\,,
\label{eq:230}
\eea
and the equalities 
\beq
\nabla^2\Phi_i(x\ta_f,z_1)\Big|_{\ta_f=0}=x^2\nabla^2\Phi_i(\ta_f,z_1)\Big|_{\ta_f=0}\,,\,\,\,\,\,
\nabla^2\Phi_i(\ta_f,z_1)\Big|_{\ta_f=0}=-\langle p_\perp^2(z_1)\rangle_0\,.
\label{eq:240}
\eeq
The terms $I_{1,2}$ in (\ref{eq:180}) arise from calculating the
Laplacian in $\ta_f$
of the first term on the right hand side of (\ref{eq:150}), and $I_3$
of the third term. The second term on the right hand side of (\ref{eq:150})
does not contribute to  $\langle p_\perp^2\rangle_r$) because
its Laplacian (for sum of the real and virtual terms) vanishes at $\ta_f=0$.
The integral over $p_\perp$ in (\ref{eq:220}) diverges logarithmically for
large  $p_{\perp}$.
In  numerical  calculations,  we regularized it by 
limiting the integration region to  
$p_\perp<p_\perp^{max}$ with $p_\perp^{max}=E\mbox{min}(x,(1-x))$.
We use expressions (\ref{eq:180})--(\ref{eq:210}) for numerical
calculations. The formulas
required for calculating $I_{1,2}$ in the oscillator approximation are given in Appendix.

The integrand in the formula (\ref{eq:190})  for $I_1$
behaves as $1/\Delta z$  for $\Delta z\to 0$, 
which leads to the logarithmic divergence of $I_1$.
In the LCPI approach it is assumed that
the typical $|z_2-z_1|$ (i.e., the formation length $L_f$)
is bigger than the correlation radius of the medium. For the
QGP it is the Debye radius. For this reason,
it is reasonable to  regularize the integration over $\Delta z$
taking the  lower  limit  in (\ref{eq:190}) at $\Delta z\sim 1/m_D$. 
This prescription was used in  \cite{Mueller_pt}
in  calculating with  a  logarithmic  accuracy the analogue of our
contribution $I_1$.
As we mentioned in the Introduction,
in \cite{Mueller_pt}
the  Glauber  factors $\Phi_{i,f}$ have not been accounted for.
For  this  reason,  the terms  
$I_{2,3}$ which contain $\nabla^2 \Phi_i$ have been missed. 
It  will  be  seen  from  the numerical  calculations
that  these terms give a negative  contribution 
which is larger  in  magnitude  than  $I_1$.

One remark should be made about the application of the above formulas
to the $p_\perp$-broadening in the real QGP. 
Formally, in the oscillator approximation, one can use a unique transport
coefficient for calculating the Green function and the Glauber factors.
However, physically it is clear that 
the values of the transport coefficient that enters the Hamiltonian
(via the oscillator frequency (\ref{eq:a70})) and the Laplacian of the
Glauber factor $\Phi_i$ (\ref{eq:240}) (via $\langle p_\perp^2\rangle_{0}$)
may differ due to the Coulomb effects.
The transport coefficient of the fast quark
(we will denote it $\hat{q}'$, and leave the
notation $\hat{q}$ for the transport coefficient that enters the
oscillator frequency)
that controls $\langle p_\perp^2\rangle_{0}$
can be written in a simple probabilistic form \cite{BDMPS2,Baier_q,JET_q}
\beq 
\hat{q}'=n\int_{0}^{p_{\perp max}^2} dp_\perp^2 p_\perp^2\frac{d\sigma}{d
  p_\perp^2}
\label{eq:250}
\eeq
with $p^2_{\perp max}\sim 3E T$, 
$T$ the QGP temperature, $d\sigma/d p_\perp^2$ the differential
cross section for quark scattering off the thermal parton.
The transport coefficient for the Hamiltonian for the
induced gluon emission
it is reasonable to define as $\hat{q}=2nC(\rho_{eff})$
\cite{LCPI1,LCPI_YF98}, where $\rho_{eff}$ is the typical size of the $gq$-state.
From the Schr\"odinger diffusion relation one obtains for
soft gluons $\rho_{eff}\sim \sqrt{2L_{f}^{eff}/\omega}$. Here
$L_f^{eff}$ is the effective in-medium gluon formation length, which
in the oscillator approximation is given by the inverse oscillator
frequency $1/|\Omega|$\footnote{It can be easily obtained from
the Schr\"odinger diffusion relation
and the relation $\hat{q}_g\rho_{eff}^2L_{f}^{eff}/4\sim 1$ (which
says that for the $qg\bar{q}$ system attenuation becomes strong at the
longitudinal scale $L_f^{eff}$). Here we used the gluon transport coefficient,
because for soft gluons $qg\bar{q}$ system interacts with the medium as
a color singlet $gg$-pair.}. For the relevant region $\rho_{eff}\lsim 1/m_D$,
with the help of the double gluon formula (\ref{eq:130}), one can show that
the product $2nC(\rho_{eff})$ coincides with (\ref{eq:250})
but with $p_{max}^2\sim
10/\rho_{eff}^2$ (see, e.g. \cite{NZ-peaks}).
This prescription gives the value of the transport coefficient
$\hat{q}$ about $\hat{q}'$ calculated for the quark energy $\sim \omega$.
For RHIC and LHC conditions,
for gluon emission from a quark with $E\sim 30-100$ GeV
the typical gluon energy $\bar{\omega}\sim 3-5$ GeV, i.e.,
$\bar{\omega}/E\ll 1$.   
Although, $\hat{q}'$ has a smooth (logarithmic) dependence on $E$,
the ratio $r=\hat{q}'/\hat{q}$
may differ significantly from unity
due to the fact that $E\gg \bar{\omega}$. As will be seen below
this gives a considerable effect
on the magnitude of $\langle p_\perp^2\rangle_r$.

Finally, we would like to make a remark on the physical interpretation
of the $I_3$ term, which,  as will be seen below, may dominate in the sum
(\ref{eq:180}).
From (\ref{eq:210}) one sees that
$I_3$ contains $\langle p_\perp^2\rangle_0$ for the whole medium,
and the vacuum spectrum $dP_v/dx$ without medium modification.
At first glance, this says that the $I_3$ is connected
with real and virtual gluon emission outside the medium from
the initial parton which has undergone multiple scattering in the whole
medium. However, this interpretation is completely wrong. In reality
the vacuum like gluon emission occurs at the longitudinal distances
which may be much smaller than the QGP size. Say, for jets with
$E\lsim 100$ GeV the typical $z$-scale for the vacuum like gluon emission
is $\lsim 1$ fm \cite{RAA08}, and the typical jet path length in
the QGP is $\sim 5$ fm. For this reason, it is clear that
typically for the vacuum like gluon emission we have a situation
with multiple scattering in the QGP
of the final partons (say, if the QGP is formed at $\tau_0\gsim 1$ fm,
the contribution of the initial parton rescatterings will be
very small). The form of the $I_3$ term is just a nontrivial
consequence of the representation  of the product
$\Phi_f\hat{g}{\cal K}\Phi_i$  in the rearranged form on the right hand
side of (\ref{eq:150}), and of the fact that the color charge of the final
quark equals to that for the initial quark. But one should bear in mind
that the decomposition on the right hand side of (\ref{eq:150}) by
adding and subtracting the terms with the vacuum Green function
is an artificial procedure, and only the full sum, given on the left hand
side of (\ref{eq:150}), matters. For this reason,
all the terms in (\ref{eq:180}) have the same status,
and it does not make sense to say that the $I_3$ is connected with a
specific mechanism due to rescatterings of the initial
quark in the whole medium and its subsequent splitting
into $gq$ state outside the QGP. Of course, such processes are possible,
but their contribution to $p_\perp$-broadening becomes very small
at $L\gg 1$ fm.


\section{ Numerical results}
We will consider $p_\perp$-broadening for conditions of
central heavy ion collisions at RHIC and LHC.
We assume that the plasma fireball is produced at the proper time
$\tau=0.5$ fm. For the fast parton path length in the QGP
we take $L=5$ fm, which is the typical jet path length in the QGP
for for Au+Au(Pb+Pb) collisions at RHIC(LHC).
We neglect the variation of the initial QGP temperature in the impact
parameter. In this case, $z$-dependence of the transport coefficient
along the fast parton path coincide with its $\tau$-dependence.
We describe the QGP evolution at $\tau>\tau_0$
within Bjorken's model \cite{Bjorken} without the transverse expansion
that leads to the entropy density $s\propto 1/\tau$.
Within the ideal gas model it gives $\hat{q}(z)=\hat{q}_0(\tau_0/z)$
at $z>\tau_0$,
where $\hat{q}_0$ is the value of the transport coefficient at $\tau=\tau_0$.
To account for qualitatively the fact that the QGP formation is not
instantaneous we take $\hat{q}(z)=\hat{q}_0(z/\tau_0)$ for
$z<\tau_0$.

For main variant we use the quasiparticle masses
$m_{q}=300$ MeV and $m_{g}=400$ MeV, which were obtained
within quasiparticle model from the  lattice  data  in \cite{LH}
for temperatures relevant for RHIC and LHC conditions.
With these values of the quasiparticle masses, in \cite{RAA13,RPP14},  we
successfully  described  the  RHIC  and  LHC  data  on the nuclear
modification  factor $R_{AA}$.
To   understand   the uncertainties  associated   with the parton  masses,
we  also perform calculations for masses $m_{q}=150$ MeV and  $m_{g}=200$ MeV.
As  in  \cite{Mueller_pt},  we take $\alpha_s=1/3$ at the vertex of 
the $q\to qg$ transition. Also, like in \cite{Mueller_pt}, we regularize 
the $1/\Delta z$ divergence in (\ref{eq:190}) by
truncating  the integration  at $\Delta z_{min}=1/m$ with $m=300$ MeV.
In (\ref{eq:190})--(\ref{eq:210})
we integrate over $x$ from $x_{min}=m_q/E$
up to $x_{max}=1-m_g/E$ (recall that we define  $x$ as $x_q$;
in terms of $x_g$, our integration region corresponds to the variation
of $x_g$ from $m_g/E$ to $1-m_q/E$). 

To fix the value of the parameter $\hat{q}_0$ in the above parametrization
of $\hat{q}(z)$ we use the results of our previous analyses of
jet quenching beyond
the oscillator approximation.
In \cite{RAA13,RPP14} we have performed calculations
of $R_{AA}$ with running $\alpha_s$ beyond the the oscillator approximation
with accurate treatment of the Coulomb effects.
To make our analysis as accurate as possible 
we adjusted the value of $\hat{q}_0$ to reproduce the quark
energy loss $\Delta E$ for the running $\alpha_s$
in the model of \cite{RPP14} with the Debye mass from the lattice
calculations \cite{Bielefeld_Md}. 
This procedure gives $\hat{q}_0\approx 0.551$ GeV$^3$ at $E=30$ GeV 
for Au+Au collisions
at $\sqrt{s}=0.2$ TeV  and 
$\hat{q}_0\approx 0.719$ GeV$^3$ at $E=100$ GeV for Pb+Pb collisions
at $\sqrt{s}=2.76$ TeV 
\footnote{We use the transport coefficient
 of the quark which is smaller than the gluon transport coefficient by a
 factor of $C_F/C_A=4/9$.}
(we call these variants the RHIC(LHC) versions). In terms of the $\hat{q}_{st}$
for the static scenario
given by
(\ref{eq:30}) our RHIC(LHC) versions correspond to $\hat{q}_{st}
\approx0.103(0.134)$ GeV$^3$. 
We have used a similar running $\alpha_s$ and the Debye mass 
to determine the introduced in section 2 the coefficient $r$
describing the enhancement of the transport coefficient entering the Glauber
factor.
We obtained for the RHIC(LHC) versions
$r\approx 2(2.3)$. We are fully aware that the errors in the factor $r$
may be rather large. But the fact that $\hat{q}'/\hat{q}$ should be
$\sim 2$ seems to be fairly reliable.
\begin{table}
  \caption{The results for $m_q=300$ MeV, $m_g=400$ MeV
    obtained with $\hat{q}'=\hat{q}$ and with $\hat{q}'=r\hat{q}$ 
    (the numbers in brackets)}
\centering
\begin{tabular}{c c c| c c}
 & \multicolumn{2}{c}{RHIC} & \multicolumn{2}{c}{LHC}\\
\hline
  & expanding & static & expanding & static \\
\hline
$I_1/\langle p_\perp^2\rangle_0$ & $0.382(0.188)$ & $0.503(0.25)$ & $0.738(0.318)$ & $0.938(0.404)$\\
\hline
$I_2/\langle p_\perp^2\rangle_0$ & $-0.277$ & $-0.212$ & $-0.145$ & $-0.113$ \\
\hline
$I_3/\langle p_\perp^2\rangle_0$ & $-0.668$  & $-0.668$ & $-1.01$ & $-1.01$\\
\hline
$\langle p_\perp^2\rangle_r/\langle p_\perp^2\rangle_0$ & $-0.563(-0.757)$
& $-0.378(-0.63)$ &
$-0.416(-0.836)$ & $-0.184(-0.719)$
\\
\hline
$\langle p_\perp^2\rangle_0$ [GeV$^2$] & $3.9(7.96)$
& $2.61(5.3)$ &
$5.1(11.85)$ & $3.4(7.89)$\\
\hline
\end{tabular}
\end{table}  

\begin{table}
  \caption{The same as in Table I, but for
    $m_q=150$ MeV, $m_g=200$ MeV}
\centering
\begin{tabular}{c c c| c c}
 & \multicolumn{2}{c}{RHIC} & \multicolumn{2}{c}{LHC}\\
\hline
  & expanding & static & expanding & static \\
\hline
$I_1/\langle p_\perp^2\rangle_0$
& $0.468(0.231)$ & $0.617(0.304)$ &$0.815(0.351)$ & $1.046(0.451)$\\
\hline
$I_2/\langle p_\perp^2\rangle_0$
& $-0.309$ & $-0.237$ & $-0.152$ & $-0.118$ \\
\hline
$I_3/\langle p_\perp^2\rangle_0$
& $-0.863$  & $-0.863$ & $-1.21$ & $-1.21$\\
\hline
$\langle p_\perp^2\rangle_r/\langle p_\perp^2\rangle_0$
& $-0.704(-0.941)$ & $-0.483(-0.796)$ &$-0.547(-1.011)$ & $-0.282(-0.877)$
\\
\hline
$\langle p_\perp^2\rangle_0$ [GeV$^2$] & $3.9(7.96)$
& $2.61(5.3)$ &
$5.1(11.85)$ & $3.4(7.89)$\\
\hline
\end{tabular}
\end{table}

In Table I we present the results for the ratio of the terms  
$I_{1,2,3}$ and of the total $\langle p_\perp^2\rangle_r$
to the $\langle p_\perp^2\rangle_{0}$ for our main
variant of the parton masses ($m_q=300$ MeV and $m_g=400$ MeV).
We also give $\langle p_\perp^2\rangle_{0}$.
We present the results both for the expanding and for static models.
For the static case the calculations are performed using $\hat{q}_{st}$. 
The results for the set $m_q=150$ MeV and $m_g=200$ MeV
are given in Table II.
The comparison of the results from Tables I and II, shows
that the sensitivity of the predictions to the parton masses turns out
to be not very strong.
Note that the sensitivity of  the induced gluon
emission  to  the  mass  of the light quark is generally low
(except for the emission of hard gluons with $x_g\sim 1$), and
the change in the predictions is mainly due to variation of $m_g$.

From Tables I and II, one can see that in all the cases
$\langle p_\perp^2\rangle_r<0$.
This occurs because
the negative contribution
from the $I_{2,3}$ terms turns out to be larger in magnitude than the positive
contribution of the $I_1$.
Note that, as we expected,
the relative effect of the negative contribution
to the mean $p_\perp^2$ from the Glauber factors
becomes bigger for the expanding QGP.
The negative $\langle p_\perp^2\rangle_r$ can lead to a sizable
reduction of the total (non-radiative
plus radiative) mean $p_\perp^2$. For the version with
$\hat{q}'=\hat{q}$ the reduction is approximately by half for the
expanding scenario. For the version with $\hat{q}'>\hat{q}$
the magnitude of the negative radiative contribution is comparable
to that from the non-radiative mechanism, and
the total mean $p_\perp^2$ may be very small.
However, one should bear in mind that this conclusion
may depend on the value of the $I_1$ term, which requires
the $\Delta z$-regularization. To understand the sensitivity of the results
to the lower limit of the $\Delta z$-integration in
(\ref{eq:190}), we have performed calculations for
$\Delta z_{min}=1/m$ with $m=600$ MeV. In this case $I_1$ becomes
bigger by a factor of $\sim 2.5(2)$ for RHIC(LHC).
This gives
$\langle p_\perp^2\rangle_r/\langle p_\perp^2\rangle_0\approx 0.01(-0.47)$
for $\hat{q}'/\hat{q}=1(r)$ for RHIC,
and
$\langle p_\perp^2\rangle_r/\langle p_\perp^2\rangle_0\approx 0.37(-0.5)$
for $\hat{q}'/\hat{q}=1(r)$  for LHC.
Thus, we see that, for the clearly more realistic version
with $\hat{q}'>\hat{q}$, the radiative correction to the mean $p_\perp^2$
is negative, and may suppress the mean $p_\perp^2$ by a factor of $\sim 2$.

In the above, we presented results of the fixed coupling
computations within the oscillator approximation for the dipole cross section.
Accurate calculations of the $I_{1,2}$ terms for the
running coupling and the double gluon $\sigma_{q\bar{q}}(\rho)$ (\ref{eq:130})
is a complicated problem. However, for the ratio
$I_3/\langle p_\perp^2\rangle_0$ the form of $\sigma_{q\bar{q}}(\rho)$
is unimportant, and the generalization to the running $\alpha_s$
can be easily done by replacing in $dP_{v}/dx$ the fixed $\alpha_s$ by
the running one \cite{Z-pt-JETP}. We performed such calculations
with the one-loop  $\alpha_s$  frozen for small momenta at the value
$\alpha_{s}^{fr}=0.7$, which was obtained earlier  from
analysis  of  the low-$x$ structure  functions within
the dipole BFKL equation \cite{NZ_HERA}. This  value is also supported
by the analysis  of heavy  quark  energy  loss  in  vacuum  \cite{DKT}.
For expanding scenario (and parton masses as in Table I),
this procedure, while keeping $I_{1,2}/\langle p_\perp^2\rangle_0$ unchanged,
 gives 
$\langle p_\perp^2\rangle_r/\langle p_\perp^2\rangle_0\approx -0.76(-0.97)$
for $\hat{q}'/\hat{q}=1(r)$ for RHIC,
and
$\langle p_\perp^2\rangle_r/\langle p_\perp^2\rangle_0\approx -0.5(-0.92)$
for $\hat{q}'/\hat{q}=1(r)$  for LHC.
These numbers are for the $\Delta z$-regularization of $I_1$ with $m=300$ MeV.
For $m=600$ MeV we obtained
$\langle p_\perp^2\rangle_r/\langle p_\perp^2\rangle_0\approx -0.19(-0.59)$
for $\hat{q}'/\hat{q}=1(r)$ for RHIC,
and
$\langle p_\perp^2\rangle_r/\langle p_\perp^2\rangle_0\approx 0.29(-0.58)$
for $\hat{q}'/\hat{q}=1(r)$  for LHC. As one can see, for the running
$\alpha_s$, the effect
of the initial state rescatterings becomes somewhat stronger.
As far as the accurate predictions for
$I_{1,2}/\langle p_\perp^2\rangle_0$ are concerned,
intuitively, one can expect that the accurate calculations
should give smaller values of $I_{1,2}/\langle p_\perp^2\rangle_0$
than obtained in the present analysis.
Indeed,
for the Green functions in the formulas (\ref{eq:190}) and (\ref{eq:200})
the typical size of the
intermediate three-body state (in the sense of their
path integral representations) should be, more or less, similar to
that for the induced gluon emission.
But we adjusted $\hat{q}$ (i.e. $2nC(\rho_{eff})$)
to reproduce the induced gluon emission energy loss obtained with
the running coupling and accurate $\sigma_{q\bar{q}}(\rho)$. 
This means that the variation of the $I_{1,2}/\langle p_\perp^2\rangle_0$
should be approximately similar to the variation of
of the ratio $C(\rho_{eff})/C(1/p_{\perp\,max})$
(because $\langle p_\perp^2\rangle_0\propto C(1/p_{\perp\,max})$.
However, for the accurate $C(\rho)$ this ratio will be smaller
than that in the oscillator approximation (see e.g. \cite{Z-rand}).
Note that this occurs even without the logarithmic growth of $C(\rho)$
at $\rho\to 0$.
Thus, one can expect that
the accurate calculations should give
smaller $I_{1,2}/\langle p_\perp^2\rangle_0$. As a results,
the effect of the negative contribution
from the $I_3$ term on the $\langle p_\perp^2\rangle_r/\langle p_\perp^2\rangle_0$
will be more pronounced.

\section{CONCLUSIONS}
We  have studied the radiative $p_\perp$-broadening  of fast  partons
in an expanding QGP for conditions of central  Au+Au(Pb+Pb) collisions at
RHIC(LHC).
The analysis  has  been  performed
within the LCPI  formalism  \cite{LCPI1,LCPI_PT}  in  the  oscillator
approximation, accounting for
the initial state rescatterings.
Similarly to the case of the static QGP,
addressed in \cite{Z-pt-JETPL,Z-pt-JETP},
we have found that the radiative correction may be negative, i.e., it
may lead to reduction of $p_\perp$-broadening.
The negative contribution to $\langle p_\perp^2\rangle_r$
comes mostly from the difference of the initial state
Glauber factors for the real and virtual processes. This
effect
appears naturally beyond the soft gluon approximation.
Formally, this phenomenon is due to
rescatterings of the initial parton for the vacuum like gluon emission.
However, we argue that this interpretation is wrong, and the effect
is dominated by rescatterings of the final fast parton.

We have found that the QGP expansion leads to a sizeable
increase of the effect of the initial state rescatterings,
as compared to the static QGP.
Our numerical results show that for the RHIC and LHC  conditions,  
due to the  negative $\langle p_\perp^2\rangle_r$
the total (non-radiative plus radiative)
mean $p_\perp^2$ may be quite small.
In light of this, it is possible that the negative experimental searches
for the jet
rescatterings in the QGP \cite{STAR1,ALICE_hjet,ALICE_hjet2}
may be due to a considerable reduction
of $p_\perp$-broadening by the radiative contribution.

\begin{acknowledgments}
  I am grateful to Peter Jacobs for drawing my attention to
  a new measurement of the jet $p_\perp$-broadening by the ALICE
  Collaboration \cite{ALICE_hjet2} and 
  helpful communication about the ALICE analysis of the
  jet deflection. 
  This work is supported by the Program 0033-2019-0005 of the Russian
  Ministry of Science and Higher Education.

\end{acknowledgments}

\appendix
\section{Formulas necessary for calculating the factors $I_i$}
In this appendix, we give formulas necessary for numerical calculations
of the contributions $I_i$ in (\ref{eq:180}) with the help
of (\ref{eq:190})--(\ref{eq:210}) in the oscillator approximation.
For quadratic parameterization of
the dipole cross section $\sigma_{q\bar{q}}(\rho)=C\rho^2$  (in terms of
quark transport coefficient $C=\hat{q}/2n$), 
the three-body cross section $\sigma_{\bar{a}bc}$
can be written as
\beq
\sigma_{\bar{a}bc}(\ro, \Rb)=
C_{b\bar{a}}(\ro_{b}-\ro_{\bar{a}})^2+C_{c\bar{a}}(\ro_{c}-\ro_{\bar{a}})^2
+C_{bc}(\ro_{b}-\ro_{c})^2\,,
\label{eq:a10}
\eeq
where, $\ro=\ro_{b}-\ro_c$, $\Rb=x_c\ro_b+x_b\ro_c-\ro_{\bar{a}}$,
$\ro_{b}-\ro_{\bar{a}}=\Rb+x_c\ro$, and $\ro_{c}-\ro_{\bar{a}}=\Rb-x_b\ro$.
For process $q\to qg$ ($a=b=q$, $c=g$) 
\beq
C_{bc}=C_{c\bar{a}}=\frac{9C}{8}\,,\,\,\,
C_{b\bar{a}}=-\frac{C}{8}\,.
\label{eq:a20}
\eeq
For diagrams in Fig.~1a, $\Rb=\ta_i$. 
It is convenient to write $\sigma_{\bar{a}bc}$ as 
\beq
\sigma_{\bar{a}bc}(\ro, \Rb)=
Cp\Rb^2+C_3\ub^2\,,
\label{eq:a30}
\eeq
where $C_3=C_{b\bar{a}}x_c^2+C_{c\bar{a}}x_b^2+C_{bc}$,
$p=[C_{b\bar{a}}C_{c\bar{a}}+C_{b\bar{a}}C_{bc}+C_{c\bar{a}}C_{bc}]/CC_3$,
and the new variable $\ub$ is given by
\beq
\ub=\ro+
\ta_iV
\label{eq:a40}
\eeq
with $V=(x_cC_{b\bar{a}}-x_bC_{c\bar{a}})/C_3$.
The Hamiltonian
(\ref{eq:100}) 
can be written in terms of the variable $\ub$ in the form
\beq
H=H_{osc}-\frac{i \hat{q}(z)p\ta_i^2}{4}+\frac{\epsilon^2}{2M}\,,
\label{eq:a50}
\eeq
where $\hat{q}(z)=2n(z)C$ is the local transport coefficient, 
and 
$H_{osc}$ is  the  harmonic oscillator
Hamiltonian
\beq
H_{osc}=
-\frac{1}{2M}\,
\left(\frac{\partial}{\partial \ub}\right)^{2}
+\frac{M\Omega^2\ub^2}{2}
\label{eq:a60}
\eeq
with the complex $z$-dependent frequency
\beq
\Omega=\sqrt{\frac{-i\hat{q}(z)C_3}{2CM}}\,.
\label{eq:a70}
\eeq   
The Green function ${\cal{K}}$ for the Hamiltonian 
(\ref{eq:a50}) can be written as
\beq
{\cal{K}}(\ro_2,z_2|\ro_1,z_1)=K_{osc}(\ub_2,z_2|\ub_1,z_1)
\exp{\left[-\frac{p\ta_i^2}{4}\int_{z_1}^{z_2}q(z)
-\frac{i(z_2-z_1)\epsilon^2}{2M}\right]}\,,
\label{eq:a80}
\eeq
where $K_{osc}$ is the Green  function  for  the oscillator  
Hamiltonian (\ref{eq:a60}).
In general, for arbitrary $\Omega(z)$ the oscillator Green function 
can be written in the form \cite{BDMS,Z_JPsi}
\beq
K_{osc}(\ub_2,z_2|\ub_1,z_1)=\frac{\gamma}{2\pi i}
\exp{\left[i(\alpha\ub_2^2+\beta \ub_1^2-\gamma\ub_1\cdot\ub_2)\right]}\,.
\label{eq:a90}
\eeq
The numerical method for evaluation of $\alpha$, $\beta$, and $\gamma$
will be discussed below.

In our formulas (\ref{eq:190})--(\ref{eq:210}), the differential operator
$\hat{g}$ (\ref{eq:110})  is
acting  on  the  Green  function  ${\cal{K}}$  at fixed
$\ta_i$. Therefore, in $\hat{g}$,
we  can  replace 
$\frac{\partial}{\partial \ro_2}\cdot\frac{\partial}{\partial \ro_1}$ by
$\frac{\partial}{\partial \ub_2}\cdot\frac{\partial}{\partial \ub_1}$.
Then, from (\ref{eq:a90}) one obtains
\beq
\frac{\partial}{\partial \ro_2}\cdot   
\frac{\partial}{\partial \ro_1}
{\cal{K}}(\ro_2,z_2|\ro_1,z_1)=
-\left[2i\gamma+
(2\alpha\ub_2-\gamma\ub_1)\cdot(2\beta\ub_1-\gamma\ub_2)\right]
{\cal{K}}(\ro_2,z_2|\ro_1,z_1)\,.
\label{eq:a100}
\eeq
For the diagram in Fig.~1a, $\ro_1=0$, $\ro_2=\ta_f=\ta_i/x_b$, 
and
\beq
\ub_{1,2}=\ta_f k_{1,2}\,,\,\,\, k_1=x_bV\,,\,\,\,
k_2=1+x_bV\,.
\label{eq:a110}
\eeq
Then from (\ref{eq:110}) 
we obtain
\beq
\hat{g}{\cal{K}}(\ro_2,z_2|\ro_1,z_1)\Big|_{\ro_{1,2}=\ta_f=0}=
-\left(\frac{\alpha_sP_{ba}}{2M^2}\right)\cdot
\frac{\gamma^2}{\pi}\exp{\left[-\frac{i(z_2-z_1)\epsilon^2}{2M}\right]}\,.
\label{eq:a120}
\eeq
For  calculating  $I_1$ (\ref{eq:190}), we need the
Laplacian in $\ta_f$ for $\ta_f=0$ of 
$\hat{g}{\cal{K}}(\ro_2,z_2|\ro_1,z_1)$ at $\ro_{2}=\ta_f$ and $\ro_1=0$.
A simple calculation gives
\beq
\nabla_{\ta_f}^2
\hat{g}{\cal{K}}(\ro_2,z_2|\ro_1,z_1)\Big|_{\ro_2=\ta_f,\ro_1=0,\ta_f=0}=
\left(\frac{\alpha_sP_{ba}}{2M^2}\right)\cdot\frac{2\gamma
(2\gamma D-G)}{i\pi}\exp{\left[-\frac{i(z_2-z_1)\epsilon^2}{2M}\right]}\,,
\label{eq:a130}
\eeq
where
\beq
D=\alpha k_2^2+\beta k_1^2-\gamma k_1k_2+
\frac{ix_bp}{4}\int_{z_1}^{z_2}dz \hat{q}(z)
\,,
\label{eq:a140}
\eeq
\beq
G=(2\alpha k_2-\gamma k_1)(2\beta k_1-\gamma k_2)\,.
\label{eq:a150}
\eeq

For the vacuum Green function, one just has to replace in 
formulas (\ref{eq:a120}) and (\ref{eq:a130})
the functions $\alpha$, $\beta$, and $\gamma$
by their vacuum analogues
\beq
\alpha_0=\beta_0=\gamma_0/2=\frac{M}{2(z_2-z_1)}\,.
\label{eq:a160}
\eeq
and to set $\hat{q}(z)=0$. In this case one obtains $D_0=\alpha_0$, $G_0=0$.

The formula (\ref{eq:a120}) holds for the Green
function $\tilde{\cal{K}}$ for the virtual diagram in Fig.~1b as well.
In the virtual counterparts of the formulas (\ref{eq:a140}) and (\ref{eq:a150}) 
$\tilde{k}_{1,2}=V$, and in the last term on 
the right-hand side of (\ref{eq:a140}) the factor $x_b$ is absent. 
For the virtual vacuum contribution 
$\tilde{D}_0=\tilde{G}_0=0$.

Let us finally discuss evaluation of the functions
$\alpha$, $\beta$, and $\gamma$.
For a harmonic oscillator with a $z$-independent frequency $\Omega$
\beq
\alpha=\beta=\frac{M\Omega}{2\tg{(\Omega(z_2-z_1))}}\,,\,\,\,\,\,
\gamma=\frac{M\Omega}{\sin{(\Omega(z_2-z_1))}}\,.
\label{eq:a170}
\eeq
For numerical calculations in the case of $z$-dependent frequency 
$\Omega(z)$ we use the $z$-slicing method based on the
recurrent relations \cite{Z_JPsi}
\bea
\alpha(z_{n+1},z_1)=\alpha(z_{n+1},z_n)-
\frac{\gamma^2(z_{n+1},z_n)}{4[\alpha(z_n,z_1)+\beta(z_{n+1},z_n)]}\,,
\nonumber\\
\beta(z_{n+1},z_1)=\beta(z_{n},z_1)-
\frac{\gamma^2(z_{n},z_1)}{4[\alpha(z_n,z_1)+\beta(z_{n+1},z_n)]}\,,
\\
\gamma(z_{n+1},z_1)=
\frac{\gamma(z_{n},z_1)\gamma(z_{n+1},z_n)}{2[\alpha(z_n,z_1)
    +\beta(z_{n+1},z_n)]}\,.
\nonumber\label{eq:a180}
\eea
These relation can be readily obtained
using (\ref{eq:a90}) and the convolution formula for the Green functions
\beq
K(\ro_3,z_3|\ro_1,z_1)=\int d\ro_2
K(\ro_3,z_3|\ro_2,z_2)
K(\ro_2,z_2|\ro_1,z_1)\,.
\label{eq:a190}
\eeq

\end{document}